\documentstyle[12pt,a4wide,cite,rotate,epsf]{article}

\newcommand{\ab}{\bar a}
\newcommand{\wh}{\widehat}
\newcommand{\gf}{\gamma_5}
\newcommand{\gl}{\gamma_E}
\newcommand{\ze}{\zeta(3)}
\newcommand{\nn}{\nonumber}

\newcommand{\as}{\alpha_{s}}

\newcommand{\Mb}{\overline M}
\newcommand{\mb}{\overline m}
\newcommand{\ve}{\varepsilon}
\newcommand{\IM}{\mbox{\rm Im}}
\newcommand{\lnp}{\big<\!\!:\!}
\newcommand{\rnp}{\!:\!\!\big>}
\newcommand{\FF}{\big<aFF\big>}
\newcommand{\eqn}[1]{(\ref{#1})}

\newcommand{\mev}{\mbox{\rm MeV}}
\newcommand{\gev}{\mbox{\rm GeV}}

\newcommand{\MSb}{{\overline{MS}}}

\newcommand{\tvs}{\vbox{\vskip 6mm}}
\newcommand{\smvs}{\vbox{\vskip 8mm}}
\newcommand{\bmvs}{\vbox{\vskip 10mm}}

\newcommand{\La}{\Lambda_{\overline{MS}}}

\newcommand{\lsim}{~{}_{\textstyle\sim}^{\textstyle <}~}
\newcommand{\newsection}[1]{\section{#1}\setcounter{equation}{0}}

\begin{document}


\begin{titlepage}
\begin{flushright}
{CERN--TH.7435/94}\\
{TUM-T31-78/94}
\end{flushright}
\vspace*{3mm}
\begin{center}
{\LARGE\sf The strange quark mass from QCD sum rules}
\vspace*{12mm}\\
{\normalsize\bf Matthias Jamin}\footnote{Address after October 1st:
Institut f\"ur Theoretische Physik, Universit\"at Heidelberg,
Philosophenweg~16, D-69120~Heidelberg}\\
{\small\sl Theory Division, CERN, CH--1211 Geneva 23, Switzerland}\\
and \\
{\normalsize\bf Manfred M\"unz}\\
{\small\sl Physik-Department, Technische Universit\"at M\"unchen}\\
{\small\sl D-85747 Garching, Germany}
\vspace{16mm}\\
{\bf Abstract}
\end{center}
\noindent
The strange quark mass is calculated from QCD sum rules for the divergence
of the vector as well as axial-vector current in the next-next-to-leading
logarithmic approximation. The determination for the divergence
of the axial-vector current is found to be unreliable due to large
uncertainties in the hadronic parametrisation of the two-point function.

{}From the sum rule for the divergence of the vector current, we obtain a value
of $m_s\equiv\mb_s(1\,\gev)=189\pm32\,\mev$, where the error
is dominated by the unknown perturbative ${\cal O}(\alpha_s^3)$
correction. Assuming a continued geometric growth of the perturbation
series, we find $m_s=178\pm18\,\mev$. Using both determinations
of $m_s$, together with quark-mass ratios from chiral perturbation
theory, we also give estimates of the light quark masses $m_u$ and $m_d$.
\vfill
\begin{flushleft}
{CERN--TH.7435/94}\\
{TUM-T31-78/94}\\
{September 1994}
\end{flushleft}

\end{titlepage}

\newpage
\setcounter{page}{1}


\newsection{Introduction}

Quark masses are amongst the fundamental set of ``a priori'' unknown
parameters in the standard model (SM) of particle physics. They are not
physical observables, as there do not exist free quarks in nature,\footnote{
except presumably for the top quark} and hence depend on the renormalization
prescription applied to the model, but have the same status as the strong
QCD coupling constant $\as$.

In this work, we shall be concerned with a determination of the strange
quark mass, $m_s$, at the next-next-to-leading order (NNLO) in the
framework of QCD sum rules \cite{svz:79}. (For an overview see
\cite{rry:85,nar:89a,shi:92}.) Already a considerable number
of determinations of the strange quark mass can be found in the literature
\cite{bnry:81,gl:82,mal:82,nprt:83,gkl:84,dom:84,rr:84,dl:85,dr:87,abb:87,nar:87,bg:88,nar:89,gen:90,dgp:91},
however partly being incompatible within their errors. For this reason,
and also because in the meantime there has been progress both on the
experimental input, as well as on the theoretical expressions, we find
it justified to reconsider the evaluation of $m_s$. In addition, there
is great interest in a precise value of $m_s$ for the calculation of
direct CP-violation in the Kaon system within the SM, because the dominant
matrix elements of four-quark operators contributing to $\ve'/\ve$ scale
like $m_s^{-2}$~\cite{bjl:93,cfmr:93}.

The basic object which is investigated in the simplest version of
QCD sum rules is the two-point function $\Psi(q^2)$ of two hadronic
currents
\begin{equation}
\Psi(q^2) \; \equiv \; i \int \! dx \, e^{iqx} \,
\big<\Omega\vert \, T\{\,j_\Gamma(x)\,j_\Gamma(0)^\dagger\}\vert\Omega\big>\,,
\label{eq:1.1}
\end{equation}
where $\Omega$ denotes the physical vacuum. To be specific, in the
rest of this work $j_\Gamma(x)$ will be the divergence of vector
and axial-vector current, $j(x)$ and $j_{5}(x)$ respectively;
\begin{equation}
j(x) \; = \; i\,(M-m):\!\bar Q(x)\,q(x)\!: \,,
\qquad
j_{5}(x) \; = \; i\,(M+m):\!\bar Q(x)\,\gf\,q(x)\!: \,,
\label{eq:1.2}
\end{equation}
with $M$ and $m$ being the masses of $Q(x)$ and $q(x)$.
Note that these currents are renormalization invariant operators.
Throughout the paper, for notational simplicity, we shall drop the
index for the pseudoscalar current, but we keep differing signs
where they appear. The upper sign will always correspond to the
divergence of the vector and the lower sign to the axial-vector current.
It should be obvious that $\Psi(q^2)$ is simply proportional to the
two-point function of scalar (pseudoscalar) currents,
\begin{equation}
\Psi(q^2) \; = \; \Big(M\mp m\Big)^2\,\Pi^{S,P}(q^2) \,,
\label{eq:1.2a}
\end{equation}
where we adopted the notation of ref.~\cite{mm:93}.

After taking two derivatives of $\Psi(q^2)$ with respect to $q^2$,
$\Psi''(q^2)$ vanishes for large $q^2$, and satisfies a dispersion
relation without subtractions (for the precise conditions see \cite{bog:58}):
\begin{equation}
\Psi''(q^2) \; \equiv \; \frac{\partial^2}{(\partial q^2)^2}\,\Psi(q^2)
\; = \; 2 \int\limits_0^\infty \frac{\rho(s)}{(s-q^2-i\ve)^3}\,ds \,,
\label{eq:1.3}
\end{equation}
where $\rho(s)$ is defined to be the spectral function corresponding
to $\Psi(s)$,
\begin{equation}
\rho(s) \; \equiv \; \frac{1}{\pi}\,\IM\,\Psi(s+i\ve) \,.
\label{eq:1.4}
\end{equation}

To suppress contributions in the dispersion integral coming from
higher excited states, it is further convenient to apply a Borel
(inverse Laplace) transformation to eq.~\eqn{eq:1.3} which leads to
\footnote{All relevant formulae for the Borel transformation are
collected in appendix~A.}
\begin{equation}
{\cal B}_u\,\Psi''(q^2) \; \equiv \; \wh\Psi''(u) \; = \;
\frac{1}{u^3} \int\limits_0^\infty e^{-s/u}\rho(s)\,ds \,,
\label{eq:1.5}
\end{equation}
or slightly rewriting eq.~\eqn{eq:1.5} we obtain
\begin{equation}
u^3\,\wh\Psi''(u) = \int\limits_0^\infty e^{-s/u}\rho(s)\,ds \,.
\label{eq:1.6}
\end{equation}
As we shall discuss in detail below, the left-hand side of this equation
is calculable in renormalization group improved perturbation theory,
if $u$ can be chosen sufficiently large. Because the Borel transformation
removes the subtraction constants in the dispersion relation and
satisfies the identity
\begin{equation}
{\cal B}_u\,\Psi(q^2) \; = \; u^2\,{\cal B}_u\,\Psi''(q^2) \,,
\label{eq:1.6a}
\end{equation}
we could have worked directly with $\Psi(q^2)$. To be able to investigate
other types of sum rules in the future, we nevertheless prefer to
express our sum rule in terms of $\Psi''(q^2)$.

Under the {\em crucial} assumption of quark-hadron duality, the right-hand
side of eq.~\eqn{eq:1.6} can be evaluated in a hadron-based picture,
still maintaining the equality, and thereby relating hadronic quantities
like masses and decay widths to the fundamental SM parameters.
Generally, however, from experiments the phenomenological spectral
function $\rho_{ph}(s)$ is only known from threshold up to some energy
$s_0$. Above this value, we shall use the perturbative expression
$\rho_{th}(s)$ also for the right-hand side. This is legitimate if
$s_0$ is large enough so that perturbation theory is applicable.
The central equation of our sum-rule analysis then is:
\begin{equation}
u^3\,\wh\Psi''_{th}(u) \;=\; \int\limits_0^{s_0} e^{-s/u}\rho_{ph}(s)
\,ds + \int\limits_{s_0}^\infty e^{-s/u}\rho_{th}(s)\,ds \,.
\label{eq:1.7}
\end{equation}
In addition, in our analysis we shall use the derivative of this equation
with respect to $u$ (``first-moment sum rule''):
\begin{equation}
u\frac{\partial}{\partial u}\biggl[\,u^3\,\wh\Psi''_{th}(u)\,\biggr]
\; = \; \int\limits_0^{s_0}\frac{s}{u}\,e^{-s/u}\rho_{ph}(s)\,ds +
\int\limits_{s_0}^\infty\frac{s}{u}\,e^{-s/u}\rho_{th}(s)\,ds \,.
\label{eq:1.8}
\end{equation}

In sect.~2, we give the expressions for the theoretical part of the
sum rules for scalar and pseudoscalar currents at the next-next-to-leading
order, and in sect. 3, the phenomenological parametrisations of the
two-point functions are discussed. Sect. 4 deals with the perturbative
continuum contribution and in sect.~5, we perform the numerical
analysis of the sum rules. Finally, in sect.~6, we compare our results
to previous determinations of $m_s$ published in the literature
and give estimates for the light quark masses $m_u$ and $m_d$.

\newsection{The theoretical two-point function}

In the framework of the operator product expansion (OPE) \cite{wil:69}
the two-point function \eqn{eq:1.1} can be expanded in inverse powers
of $Q^2\equiv-q^2$:
\begin{equation}
\Psi(Q^2) \; = \; (M\mp m)^2 Q^2\,\biggl\{\,\Psi_0 + \Psi_2\frac{1}{Q^2} +
\Psi_4\frac{1}{Q^4} + \Psi_6\frac{1}{Q^6} + \ldots\,\biggr\} \,.
\label{eq:2.1}
\end{equation}
The $\Psi_n$ contain operators of dimension $n$, and their remaining
$Q^2$-dependence is only logarithmic. In this work, we shall
treat mass like an operator of dimension 1. Otherwise, the contribution
$\Psi_2$ would be absent because one cannot construct a gauge-invariant
operator of dimension 2. Likewise, $\Psi''(Q^2)$ has the expansion
\begin{equation}
\Psi''(Q^2) \; = \; \frac{(M\mp m)^2}{Q^2}\,\biggl\{\,\Psi_0'' +
\Psi_2''\frac{1}{Q^2} + \Psi_4''\frac{1}{Q^4} + \Psi_6''\frac{1}{Q^6} +
\ldots\,\biggr\} \,.
\label{eq:2.2}
\end{equation}
In the following sections, we shall calculate these expressions explicitly.

\subsection{The perturbative contribution}

In general the perturbative expression for $\Psi_0(Q^2)$ is given by
\begin{equation}
\Psi_0(Q^2) \; = \; \frac{N}{8\pi^2}\sum_{i=0}^{\infty} a^i
\sum_{j=0}^{i+1} c_{ij}\,L^j \,,
\label{eq:2.3}
\end{equation}
where $N$ is the number of colours, $a\equiv\as/\pi$ and
$L\equiv\ln(Q^2/\mu^2)$. $\mu^2$ is a renormalization scale. Throughout
this work, we shall use the modified minimal subtraction scheme
$\MSb$~\cite{bbdm:78}. From eq.~\eqn{eq:2.3}, we easily obtain the
corresponding expression for $\Psi_0''(Q^2)$:
\begin{equation}
\Psi_0''(Q^2) \; = \; \frac{N}{8\pi^2}\sum_{i=0}^{\infty} a^i
\sum_{j=1}^{i+1} c_{ij}\,j\Big[\,L^{j-1}+(j-1)L^{j-2}\,\Big] \,.
\label{eq:2.4}
\end{equation}
Let us remark that the constant coefficients $c_{i0}$ have dropped
out of this expression.

Because $\Psi''(Q^2)$ is related to a physical quantity --- the spectral
function $\rho(s)$ --- it has to satisfy a homogeneous renormalization
group equation (RGE):
\begin{equation}
\biggl\{\,-2\frac{\partial}{\partial L}+\beta(a) a\frac{\partial}{\partial a}
-\gamma(a)\biggl(M\frac{\partial}{\partial M}+m\frac{\partial}{\partial m}\,
\biggr)\,\biggr\}(M\mp m)^2\Psi_0''(Q^2) \; = \; 0 \,,
\label{eq:2.5}
\end{equation}
where $\beta(a)$ is the QCD $\beta$-function and $\gamma(a)$ is the mass
anomalous dimension. In appendix~B, we have collected the coefficients
of $\beta(a)$ and $\gamma(a)$ in our notation. In eq.~\eqn{eq:2.5}, all
coefficients multiplying a certain term $a^i L^j$ have to vanish separately.
{}From this condition one derives relations between the various $c_{ij}$
which are also given in appendix~B. The independent coefficients are
conveniently chosen to be the $c_{i1}$. The coefficients $c_{01}$
and $c_{11}$ can be straightforwardly calculated and are found to be
\begin{equation}
c_{01} \; = \; 1 \qquad \hbox{and} \qquad
c_{11} \; = \; \frac{17}{4}\,C_F \,,
\label{eq:2.6}
\end{equation}
with $C_F=(N^2-1)/2N$. The NNLO coefficient $c_{21}$ has been calculated
recently in ref.~\cite{gkls:90}. The result is
\begin{eqnarray}
c_{21} & = & C_F \biggl[\,\biggl(\frac{691}{64}-\frac{9}{4}\ze\biggr)C_F
+\biggl(\frac{893}{64}-\frac{31}{8}\ze\biggr)N-\biggl(\frac{65}{32}-
\frac{1}{2}\ze\biggr)f\,\biggr] \,, \nn \\
\smvs
& \stackrel{N=3}{=} & \frac{10801}{144}-\frac{39}{2}\ze-\biggl(
\frac{65}{24}-\frac{2}{3}\ze\biggr)f \,,
\label{eq:2.7}
\end{eqnarray}
where $f$ is the number of flavours, and $\zeta(z)$ is the Riemann
$\zeta$-function. Since $\Psi''$ satisfies a homogeneous RGE, we can
sum up the logarithms by choosing $\mu^2=Q^2$. Then the coupling and
the masses become running quantities, evaluated at $Q^2$. To NNLO and
$N=f=3$ this yields for $\Psi_0''$:
\begin{equation}
\Psi_0'' \; = \; \frac{N}{8\pi^2}\,\biggl[\,1+\frac{11}{3}\,\ab(Q^2)+
\biggl(\frac{5071}{144}-\frac{35}{2}\ze\biggr)\,\ab(Q^2)^2+
{\cal O}(\ab^3)\,\biggr] \,.
\label{eq:2.8}
\end{equation}

The final step in the evaluation of the perturbative contribution to
the two-point function consists in performing the Borel transformation.
Unfortunately, the running coupling $\ab(Q^2)$ as well as the running
masses $\Mb(Q^2)$ and $\mb(Q^2)$ at NNLO are now complicated
functions of $Q^2$, which are conventionally expanded in powers of
$\ln^{-1}(Q^2/\Lambda^2)$. The Borel transform of the resulting
expressions cannot be given in closed form, but has to be expanded
in powers of $\ln^{-1}(u/\Lambda^2)$, or calculated numerically
\cite{aya:91,aya:92}.

There is a different way to obtain the Borel transform of the two-point
function. Because the differential operators for the RGE and
the Borel transform act on different variables, $\mu^2$ and $Q^2$
respectively, they commute \cite{cdks:84}. We thus apply
the Borel operator directly to eq.~\eqn{eq:2.4}. The
necessary formulae are given in appendix~A. One can then easily
convince oneself that the resulting expression for
$(M\mp m)^2\wh\Psi_0''(u)$ again satisfies a homogeneous RGE, and
the logarithms can be summed up through the choice $\mu^2=u$.
This results in the running coupling and masses being evaluated
at the scale $u$. Our final expression for $\wh\Psi_0''(u)$ is
\begin{eqnarray}
\label{eq:2.9}
\wh\Psi_0''(u) & = & \frac{N}{8\pi^2}\,\biggl[\,1+\ab(u)\Big(c_{11}+
2(1-\gl)c_{12}\Big) \nn \\
\smvs
& & +\,\ab^2\Big(c_{21}+2(1-\gl)c_{22}+(3\gl^2-6\gl-\frac{\pi^2}{2})c_{23}
\Big) \\
\smvs & & \hspace{-2.2cm}
    +\,\ab^3\Big(c_{31}+2(1-\gl)c_{32}+(3\gl^2-6\gl-\frac{\pi^2}{2})c_{33}
-2(2\gl^3-6\gl^2-\gl\pi^2+\pi^2+4\ze)c_{34}\Big)\,\biggr] \,, \nn
\end{eqnarray}
where $\gl$ is the Euler constant. Here, analogously to eqs.~\eqn{eq:2.1}
and \eqn{eq:2.2}, $\wh\Psi''(u)$ has been expanded in inverse powers of $u$:
\begin{equation}
\wh\Psi''(u) \; = \; \frac{(\Mb(u)\mp \mb(u))^2}{u}\,\biggl\{\,\wh\Psi_0''+
\wh\Psi_2''\frac{1}{u} + \wh\Psi_4''\frac{1}{u^2} +
\wh\Psi_6''\frac{1}{u^3} + \ldots\,\biggr\} \,.
\label{eq:2.10}
\end{equation}
In our numerical analysis we have verified explicitly that our
treatment of the Borel transform leads to the same result as the method
of refs.~\cite{aya:91,aya:92}, up to differences which are of higher order.

In addition to eq.~\eqn{eq:2.9}, for the first-moment sum rule
of eq.~\eqn{eq:1.8} it is convenient to define the function
\begin{equation}
\wh\Phi(u) \; \equiv \; u\,\frac{\partial}{\partial u}\,\wh\Psi''(u) \,.
\label{eq:2.11}
\end{equation}
Expanding this function equivalently to $\wh\Psi''(u)$, eq.~\eqn{eq:2.10},
we obtain
\begin{eqnarray}
\label{eq:2.12}
\wh\Phi_0(u) & = & -\,\frac{N}{8\pi^2}\,\biggl[\,1+\ab(u)\Big(c_{11}-
2\gl c_{12}\Big)+\ab^2\Big(c_{21}-2\gl c_{22}+(3\gl^2-6-\frac{\pi^2}{2})
c_{23}\Big) \qquad \\
\smvs
& + & \,\ab^3\Big(c_{31}-2\gl c_{32}+(3\gl^2-6-\frac{\pi^2}{2})c_{33}
-2(2\gl^3-12\gl-\gl\pi^2+4\ze)c_{34}\Big)\,\biggr] \,. \nn
\end{eqnarray}

To be able to estimate the uncertainty coming from the as yet unknown
${\cal O}(a^3)$ term in the perturbative result, we have included this
contribution which depends on the coefficient $c_{31}$ in the general
expressions of eqs.~\eqn{eq:2.9} and \eqn{eq:2.12}, and in our numerical
analysis we varied this coefficient in order to simulate a
contribution from higher orders. Let us point out
however, that the ${\cal O}(a^3)$ correction cannot be included in a
completely consistent way, because this would also require the four-loop
coefficients of the $\beta$-function and the mass anomalous dimension,
which have not yet been calculated.

\subsection{Dimension 2 operators}

The contribution to the dimension 2 operators can be obtained by expanding
the perturbative result for the vector-current two-point function by
Generalis~\cite{gen:90a} in powers of the quark masses. Then the expression
for the scalar (pseudoscalar) two-point function is calculable from the
Ward-identity between the vector (axial-vector) and scalar (pseudoscalar)
two-point functions (see e.g. eq.~(2.3) of ref.~\cite{mm:93}). There is,
however, a subtlety, because this Ward-identity involves the renormalized,
{\em non}-normal-ordered quark condensate $\big<\bar\psi\psi\big>$. Since
from a straightforward application of perturbation theory we get expressions
containing the normal-ordered condensate $\lnp\bar\psi\psi\rnp$, we still
have to subtract from Generalis result a contribution which stems from
the relation between $\big<\bar\psi\psi\big>$ and $\lnp\bar\psi\psi\rnp$
\cite{sc:88}. This additional contribution removes mass logarithms
which are present at intermediate steps of the calculation \cite{bg:88,mm:93}.
For $\Psi_2$ we find
\begin{eqnarray}
\Psi_2(Q^2) & = & \frac{N}{8\pi^2}\,\biggl\{\,\biggl[\,2L-2+aC_F\Big(
-3L^2+8L-\frac{25}{2}+6\ze\Big)\,\biggr]\Big(M^2+m^2\Big) \nn \\
\smvs
& & \hspace{7.3mm} \pm\,\biggl[\,2L-4+aC_F\Big(-3L^2+14L-22+6\ze\Big)\,
\biggr]\,Mm\,\biggr\} \,.
\label{eq:2.13}
\end{eqnarray}
This result is in agreement with Chetyrkin et al. \cite{cgs:85}, who
also performed a calculation for the vector current, but already
gave expressions in terms of {\em non}-normal-ordered condensates.

{}From eq.~\eqn{eq:2.13} we obtain
\begin{equation}
\Psi_2''(Q^2) \; = \; -\,\frac{N}{4\pi^2}\,\biggl\{\,\biggl[\,1+aC_F\Big(
-3L+7\Big)\,\biggr]\Big(M^2+m^2\Big)\pm\biggl[\,1+aC_F\Big(-3L+10\Big)\,
\biggr]\,Mm\,\biggr\} \,.
\label{eq:2.14}
\end{equation}
Proceeding for the Borel transform as in section 2.1, we finally find
\begin{equation}
\wh\Psi_2''(u) \; = \; -\,\frac{N}{4\pi^2}\,\biggl\{\,\biggl[\,1+\ab C_F\Big(
4+3\gl\Big)\,\biggr]\Big(\Mb^2+\mb^2\Big)
\pm\biggl[\,1+\ab C_F\Big(7+3\gl\Big)\,\biggr]\,\Mb\,\mb\,\biggr\} \,,
\label{eq:2.15}
\end{equation}
and
\begin{equation}
\wh\Phi_2(u) \; = \; \phantom{-}\,\frac{N}{2\pi^2}\,\biggl\{\,\biggl[\,1+\ab
C_F\biggl(\frac{11}{2}+3\gl\biggr)\,\biggr]\Big(\Mb^2+\mb^2\Big)
\pm\biggl[\,1+\ab C_F\biggl(\frac{17}{2}+3\gl\biggr)\,\biggr]\,
\Mb\,\mb\,\biggr\} \,.
\label{eq:2.16}
\end{equation}
Again, $\wh\Psi_2''(u)$ satisfies a homogeneous RGE which justifies
the exchange of Borel transformation and renormalization group summation.

\subsection{Dimension 4 operators}

There are three sources of dimension 4 operators contributing to
$\Psi(Q^2)$: explicit mass corrections $\sim m^4$, terms originating
from the quark condensate $\sim m\big<\bar\psi\psi\big>$ and the
gluon condensate contribution $\sim\FF$. To the order
considered,\footnote{For our counting of orders in perturbation
theory see the remarks in appendix~B.}
the explicit quark mass as well as the gluon condensate contribution
can, for example, be taken from ref.~\cite{mm:93}. As for the quark
condensate, to NNLO we also need ${\cal O}(\as)$ corrections. These
can either be calculated directly~\cite{pr:82}, or inferred from the
cancellation of mass logarithms after expressing $\lnp\bar\psi\psi\rnp$
through $\big<\bar\psi\psi\big>$~\cite{bg:88}, yielding identical
results. The explicit expressions are:
\begin{eqnarray}
\label{eq:2.17}
\Psi_{4,m^4} & = & \frac{N}{16\pi^2}\,\biggl[\,M^4+4M^2 m^2+m^4+
2\Big(M^4\pm2M^3 m\pm2Mm^3+m^4\Big)L\,\biggr] \,, \\
\smvs
\Psi_{4,\bar\psi\psi} & = & \frac{1}{2}\,\biggl[\,1+aC_F\biggl(-\frac{3}{2}
L+\frac{11}{4}\biggr)\,\biggr]\Big(\,M\big<\bar QQ\big>+m\big<\bar qq\big>
\,\Big) \nn \\
\smvs
& & \pm\,\biggl[\,1+aC_F\biggl(-\frac{3}{2} L+\frac{7}{2}\biggr)\,\biggr]
\Big(\,m\big<\bar QQ\big>+M\big<\bar qq\big>\,\Big) \,, \\
\smvs
\Psi_{4,FF} & = & \frac{1}{8}\,\FF \,.
\end{eqnarray}
{}From these expressions one can immediately obtain $\Psi_4''$.
It is an instructive exercise to show that again $(M\mp m)^2\Psi_4''$
does satisfy the homogeneous RGE of eq.~\eqn{eq:2.5}. To this end, the
condensates are conveniently rewritten into renormalization group
invariant condensates~\cite{nt:83,sc:88}. The formulae for the
renormalization group invariant condensates are collected in appendix~C.
In addition, to show this one needs the terms of order $\as m^4$
which can also be calculated via expanding the result by Generalis
\cite{gen:90a}.

As before, the next steps are performing the Borel transformation
and summing up the logarithms through the choice $\mu^2=u$.
This leads to
\begin{eqnarray}
\wh\Psi_4''(u) & = & \frac{1}{2}\,\Biggl\{\,\frac{1}{4}\,\Omega_4 +
\frac{\gamma_1}{\beta_1}\,\ab\sum_i \Omega^{m_iq_i}_3 - \frac{w_1}{4}
\sum_i \mb_i^4 \nn \\
\smvs
& + & \biggl[\,1+\ab C_F\biggl(\frac{11}{4}+\frac{3}{2}\gl\biggr)\,\biggr]
\Big(\Omega^{MQ}_3+\Omega^{mq}_3\Big)\pm\biggl[\,2+\ab C_F\Big(7+3\gl\Big)
\,\biggr]\Big(\Omega^{mQ}_3+\Omega^{Mq}_3\Big) \nn \\
\smvs
& - &  \biggl\{\, \biggl[\,\frac{w_1}{\ab}+w_1\biggl(C_F\biggl(\frac{11}{4}+
\frac{3}{2}\gl\biggr)+w_2\biggr)-\frac{\gamma_1^0}{4}\Big(1-2\gl\Big)\,
\biggr]\Big(\Mb^4+\mb^4\Big) - \gamma_1^0\Mb^2\mb^2 \nn \\
\smvs
& & \hspace{-0.6mm}\pm\,\biggl[\,2\frac{w_1}{\ab}+w_1\biggl(C_F\Big(
7+3\gl\Big)+2w_2\biggr)+\gamma_1^0\gl\,\biggr]\Big(\Mb^3\mb+\Mb\mb^3\Big)
\,\biggr\}\,\Biggr\} \,,
\label{eq:2.20}
\end{eqnarray}
and
\begin{eqnarray}
\wh\Phi_4(u) & = & -\,\frac{3}{2}\,\Biggl\{\,\frac{1}{4}\,\Omega_4 +
\frac{\gamma_1}{\beta_1}\,\ab\sum_i \Omega^{m_iq_i}_3 - \frac{w_1}{4}
\sum_i \mb_i^4 \nn \\
\smvs
& + & \biggl[\,1+\ab C_F\biggl(\frac{13}{4}+\frac{3}{2}\gl\biggr)\,\biggr]
\Big(\Omega^{MQ}_3+\Omega^{mq}_3\Big)\pm\biggl[\,2+\ab C_F\Big(8+3\gl\Big)
\,\biggr]\Big(\Omega^{mQ}_3+\Omega^{Mq}_3\Big) \nn \\
\smvs
& - &  \biggl\{\, \biggl[\,\frac{w_1}{\ab}+w_1\biggl(C_F\biggl(\frac{13}{4}+
\frac{3}{2}\gl\biggr)+w_2\biggr)-\frac{\gamma_1^0}{4}\biggl(\frac{1}{3}-2\gl
\biggr)\,\biggr]\Big(\Mb^4+\mb^4\Big) - \gamma_1^0\Mb^2\mb^2 \nn \\
\smvs
& & \hspace{-0.6mm}\pm\,\biggl[\,2\frac{w_1}{\ab}+w_1\biggl(C_F\Big(
8+3\gl\Big)+2w_2\biggr)+\gamma_1^0\biggl(\frac{1}{3}+\gl\biggr)\,
\biggr]\Big(\Mb^3\mb+\Mb\mb^3\Big) \,\biggr\}\,\Biggr\}
\label{eq:2.21}
\end{eqnarray}
with $\Omega_3$ and $\Omega_4$ being the renormalization group invariant
condensates as defined in appendix~C. All other definitions can be found
in appendices~B and C.

\subsection{Dimension 6 operators}

For the dimension 6 operators, we take into account only the most
important contributions. In addition, since until today the renormalization
group behaviour of dimension 6 operators has not been exploited completely,
we shall neglect the running of those operators. Anyhow, numerically their
contribution is only a very small correction, due to the suppression by
powers of $Q^2$. This will be discussed further in our numerical analysis
of section~5.

Including operators with up to one power in the quark masses, we have:
\begin{eqnarray}
\Psi_6(Q^2) & = & \pm\,\frac{1}{2}\,\Big[\,M\langle g\bar q\sigma Fq\rangle
+m\langle g\bar Q\sigma FQ\rangle\,\Big] \pm 4\pi^2 \ab\langle\bar Q
\sigma_{\mu\nu}t^a q\,\bar q\sigma^{\mu\nu}t^a Q\rangle \nn \\
\smvs
& & +\,\frac{4}{3}\,\pi^2 \ab\langle\bar Q\gamma_\mu t^a Q+\bar q
\gamma_\mu t^a q\!\!\sum_{A=u,d,s}\!\!\bar q^A\gamma^\mu t^a q^A\rangle \,.
\label{eq:2.22}
\end{eqnarray}
Here, $\langle g\bar\psi\sigma F\psi\rangle$ is the so called ``mixed''
condensate, and the other two terms are four-quark operators. Since there
exist no reliable estimates for the vacuum expectation values of these
four-quark operators in the literature, we follow the usual procedure in
sum rule analyses by using the vacuum dominance hypothesis \cite{svz:79}
to relate them to the quark condensate. After taking two derivatives and
performing the Borel transformation this leads to
\begin{equation}
\hat\Psi_6''(u) \; = \; \pm\,\frac{1}{2}\,\Big[\,M\langle g\bar q\sigma
Fq\rangle+m\langle g\bar Q\sigma FQ\rangle\,\Big]-\frac{4C_F}{3N}\,\pi^2 \ab
\Big[\,\langle\bar QQ\rangle^2+\langle\bar qq\rangle^2\pm9\langle\bar QQ
\rangle\langle\bar qq\rangle\,\Big] \,.
\label{eq:2.23}
\end{equation}
The corresponding quantity $\hat\Phi_6$ for the first moment sum rule
is simply given by $-4\hat\Psi_6''$.

\newsection{Phenomenological two-point function}

In this section, we have to distinguish between the scalar and the
pseudoscalar two-point function. Let us therefore discuss both separately.
In addition, we now have to specify the actual flavour content of the
currents of eq.~\eqn{eq:1.2}. Since in the end we want to calculate the
strange quark mass, $Q$ should be chosen to be the strange quark and
$M=m_s$. The other quark is conveniently chosen to be a light quark,
because for these channels we have reasonably good information on the
spectral functions. For definiteness, we shall use the up-quark and
$m=m_u$, neglecting isospin-breaking effects.

\subsection{The scalar two-point function}

Generally, the phenomenological spectral function $\rho(s)$ is given by
\begin{equation}
\rho_\Gamma(s) \; = \; (2\pi)^3\sum_\Gamma\hspace{-5mm}\int\;\,\langle
\Omega|j_\Gamma(0)|\Gamma\rangle\langle\Gamma|j_\Gamma(0)^\dagger|\Omega
\rangle\,\delta^4(q-p_\Gamma) \,,
\label{eq:3.0}
\end{equation}
where $\Gamma$ are intermediate states with the correct quantum numbers
over which we have to sum and calculate the corresponding phase-space
integrals.

For the scalar two-point function, the lowest lying state which contributes
to the spectral function is the $K\pi$-system in an $s$-wave $I=1/2$
state. The contribution of this intermediate state yields the inequality
\cite{nprt:83}
\begin{equation}
\rho_{ph}(s) \; \ge \; \theta(s-s_+)\,\frac{3}{32\pi^2}\,\frac{1}{s}
\,\sqrt{(s-s_+)(s-s_-)} \,\big|d(s)\big|^2 \,,
\label{eq:3.1}
\end{equation}
where
\begin{equation}
s_+ \; = \; (M_{K^+}+M_{\pi^0})^2 \,,
\qquad
s_- \; = \; (M_{K^+}-M_{\pi^0})^2 \,,
\label{eq:3.2}
\end{equation}
and $d(s)$ is the strangeness changing scalar form factor which appears
in $K_{l3}$ decays:
\begin{equation}
\langle\pi^0(p')\vert\,(\bar s\,\gamma_\mu u)(0)\vert K^+(p)\rangle \; = \;
\frac{1}{\sqrt{2}}\,\Big[\,(p+p')_\mu f_+(s)+(p-p')_\mu f_-(s)\,\Big] \,,
\label{eq:3.3}
\end{equation}
and
\begin{equation}
d(s) \; = \; (M_K^2-M_\pi^2) f_+(s)+s f_-(s) \,.
\label{eq:3.4}
\end{equation}
The physical region for $K_{l3}$ decays is $m_l^2\le s\le s_-$, whereas
in the spectral function $s\ge s_+$. In this region, $d(s)$ is not
directly accessible to experiment.

In the following, we assume that the spectral function is saturated by
a sum of Breit-Wigner resonances multiplied by the threshold behaviour
of eq.~\eqn{eq:3.1}. This leads to \cite{dl:85}
\begin{equation}
\rho_{ph}(s) \; = \; \theta(s-s_+)\,\frac{3}{32\pi^2}\,\frac{1}{s}
\,\sqrt{(s-s_+)(s-s_-)} \,\big|d(s_+)\big|^2 \,
\frac{\sum_n BW(s,F_n,M_n,\Gamma_n)}{\sum_n BW(s_+,F_n,M_n,\Gamma_n)} \,,
\label{eq:3.5}
\end{equation}
with
\begin{equation}
BW(s,F_n,M_n,\Gamma_n) \; = \; \frac{F_n^2 M_n^5 \Gamma_n}
{(M_n^2-s)^2+M_n^2 \Gamma_n^2(s)} \,.
\label{eq:3.6}
\end{equation}
$M_n$, $\Gamma_n$ and $F_n$ are the mass, width and decay constant of the
n-th resonance respectively. In the denominator we use an energy-dependent
width $\Gamma_n(s)$:
\begin{equation}
\Gamma_n(s) \; = \; \frac{M_n}{\sqrt{s}}\,\frac{q(s)}{q(M_n^2)}\,\Gamma_n\,,
\qquad \hbox{where} \qquad
q(s) \; = \; \frac{1}{2\sqrt{s}}\,\sqrt{(s-s_+)(s-s_-)}
\label{eq:3.7}
\end{equation}
is the momentum in the centre-of-mass system. In the numerator, this
energy dependence reproduces to the correct threshold behaviour of
the spectral function and has been pulled out in writing eq.~\eqn{eq:3.5}.
Experimentally, the first two resonances, namely the $K_0^*(1430)$
and the $K_0^*(1950)$, are known \cite{ast:88}, and will be taken
into account in our numerical analysis. All values for masses, widths
and other parameters are given explicitly in section~5.

The last quantity being required as an input for the phenomenological
two-point function is the scalar form factor at threshold, $d(s_+)$.
It can be calculated from an Omn\`es representation for $d(s)$ \cite{nprt:83}
\begin{equation}
d(s_+) \; = \; d(0) \exp\Biggl\{\,\frac{s_+}{\pi}\int\limits_{s_+}^\infty
\frac{\delta_1(s')}{s'(s'-s_+-i\ve)}\,ds'\,\Biggr\} \,,
\label{eq:3.8}
\end{equation}
where $\delta_1(s)$ is the $K\pi$ $s$-wave, $I=1/2$ phase shift, which
can be taken from experiment \cite{ast:88,awa:86,est:78}. In writing
eq.~\eqn{eq:3.8} we have assumed purely elastic scattering. This is
justified in the region close to threshold which dominates the integral.
In section~5, we shall also compare with other methods to obtain $d(s_+)$.

\subsection{The pseudoscalar two-point function}

In the pseudoscalar channel the lowest lying state with the relevant
flavour quantum numbers which contributes to the spectral function,
is the $K$-meson. In order to obtain good stability in the sum rule,
we shall also include the next two resonances, for which there exists
experimental evidence: the $K(1460)$ and the $K(1830)$. For the
$K$-meson, a $\delta$-resonance approximation is sufficient, because
it is relatively long living, but for the two higher resonances we
again have to use an Breit-Wigner Ansatz with a finite width. As in
the case of the scalar two-point function, we shall impose the correct
threshold behaviour on the Breit-Wigner Ansatz.

The next-higher intermediate state above the $K$-meson is the
$(K\pi\pi)$-system in an $s$-wave $I=1/2$ state. Calculating the
corresponding matrix elements for the divergence of the axial-vector
current to leading order in chiral perturbation theory, we obtain
\begin{eqnarray}
\label{eq:3.10}
\langle\Omega|\partial^\mu(\bar s\gamma_\mu\gamma_5u)(0)|K^+(p_1)
\pi^+(p_2)\pi^-(p_3)\rangle & = &  \\
\smvs & & \hspace{-8cm}
= \; -\,\frac{M_K^2}{3F_\pi}\,\biggl\{\,2+\frac{1}{M_K^2-q^2}\,\Big[
(q+p_1)(p_2-p_3)+(q+p_2)(p_1-p_3)-(M_K^2+M_\pi^2)\,\Big]\,\biggr\} \,, \nn \\
\bmvs
\label{eq:3.11}
\langle\Omega|\partial^\mu(\bar s\gamma_\mu\gamma_5u)(0)|K^+(p_1)
\pi^0(p_2)\pi^0(p_3)\rangle & = &  \\
\smvs & & \hspace{-8cm}
= \; -\,\frac{M_K^2}{6F_\pi}\,\biggl\{\,4+\frac{1}{M_K^2-q^2}\,\Big[
(q+p_2)(p_1-p_3)+(q+p_3)(p_1-p_2)-2(M_K^2+M_\pi^2)\,\Big]\,\biggr\} \,. \nn
\end{eqnarray}
In our conventions $F_\pi=132\,{\rm \mev}$.

Performing the phase-space integration, and modulating the threshold
behaviour of the higher states with a sum of Breit-Wigner resonances,
in the chiral limit the spectral function is found to be
\begin{equation}
\rho_{5\,ph}(s) \; = \; F_K^2 M_K^4\,\Biggl\{\,\delta(s-M_K^2)+
\frac{23}{3\!\cdot\!2^{12}\pi^4}\,\frac{s}{F_\pi^4}\,\frac{\sum_n
BW(s,F_n,M_n,\Gamma_n)}{\sum_n BW(0,F_n,M_n,\Gamma_n)}\,\Biggr\} \,.
\label{eq:3.12}
\end{equation}
The Breit-Wigner function is given in eq.~\eqn{eq:3.6}. Notice that
to leading order in the chiral expansion the threshold is shifted
to $s=0$. In this limit the width of the resonance becomes independent
of the energy and, therefore, for consistency, we set $\Gamma_n(s)=\Gamma_n$.

\newsection{The perturbative continuum}

Above the energy $s_0$, up to which experimental information on the
spectral function is available, we approximate the remaining
contribution by the perturbative continuum, neglecting all power
corrections, being negligible for energies greater than $s_0$.

The two integrals which have to be calculated are
\begin{equation}
\label{eq:4.2}
I_0(u) \; \equiv \; \int\limits_{s_0}^\infty e^{-s/u}\rho_{th}(s)\,ds
\qquad \hbox{and} \qquad
I_1(u) \; \equiv \; \int\limits_{s_0}^\infty\frac{s}{u}\,e^{-s/u}
\rho_{th}(s)\,ds\,.
\label{eq:4.1}
\end{equation}
The expression for $\rho_{th}(s)$ can be obtained from eq.~\eqn{eq:2.3}.
Up to ${\cal O}(a^3)$, we find
\begin{eqnarray}
\rho_{th}(s) & = & \frac{N}{8\pi^2}\,(M\mp m)^2 s\,\biggl\{\,1+a\,\Big(2c_{12}L
+c_{11}\Big)+a^2\Big(3c_{23} L^2+2c_{22} L+c_{21}-c_{23}\pi^2\Big) \nn \\
\smvs
& & +\,a^3\Big[\,4c_{34} L^3+3c_{33} L^2+(2c_{32}-4c_{34}\pi^2)L+
c_{31}-c_{33}\pi^2\,\Big]\,\biggr\} \,,
\end{eqnarray}
where $L\equiv \ln(s/\mu^2)$. Using the relations amongst the $c_{ij}$
of eqs.~\eqn{eq:b.8} and \eqn{eq:b.9}, one can convince oneself that
also $\rho_{th}(s)$ satisfies a homogeneous RGE, as it should.

Hence, to calculate $I_0(u)$ and $I_1(u)$, we need the following type
of integrals:
\begin{equation}
I(\alpha,n) \; \equiv \; \int\limits_{s_0}^\infty ds\,s^{\alpha-1}
\ln^n\frac{s}{\mu^2}\,e^{-s/u}
\; = \; u^\alpha \sum_{k=0}^n {n\choose k}\ln^k\frac{u}{\mu^2}\,
\Biggl[\,\frac{\partial^{\,n-k}}{(\partial\alpha)^{n-k}}\,
\Gamma(\alpha,y)\,\Biggr] \,,
\label{eq:4.3}
\end{equation}
with $y\equiv s_0/u$, and $\Gamma(\alpha,y)$ is the incomplete
$\Gamma$-function~\cite{gr:80}. Using this formula, we obtain:
\begin{eqnarray}
\label{eq:4.4}
I_0(u) & = & \frac{N}{8\pi^2}\,(\Mb\mp\mb)^2 u^2\,\biggl\{\,
\Gamma(2,y)\Big[\,1+\ab c_{11}+\ab^2(c_{21}-c_{23}\pi^2)+\ab^3(c_{31}-
c_{33}\pi^2)\,\Big] \\
\smvs & & \hspace{-20mm}
+\,2\Gamma'(2,y)\Big[\,\ab c_{12}+\ab^2 c_{22}+\ab^3(c_{32}-2c_{34}
\pi^2)\,\Big] + 3\Gamma''(2,y)\Big[\,\ab^2 c_{23}+\ab^3 c_{33}\,\Big] +
4\Gamma'''(2,y)\ab^3 c_{34}\,\biggr\} \nn \,.
\end{eqnarray}
The logarithms have again been summed up leading to running quantities
evaluated at $\mu^2=u$. The corresponding expression for $I_1(u)$ can
be obtained through replacing $\Gamma^{(n)}(2,y)$ by $\Gamma^{(n)}(3,y)$
in eq.~\eqn{eq:4.4}.

\newsection{Numerical Results}

\subsection{Scalar two-point function}

In our numerical analysis of the sum rules, we shall mainly discuss the
values of our input parameters, their errors, and the impact of those
errors on the strange quark mass obtained.
  In the OPE, we generically have to evaluate expressions of the type
$\ab^m(u)\Mb^{\,n}(u)$, where $m$ and $n$ are integer exponents. To achieve
a systematic expansion in perturbation theory, we have decided to
expand these terms consistently in inverse powers of $L\equiv\ln(u/\La^2)$.
The relevant formula is given in eq.~\eqn{eq:b.6} in appendix~B. Besides the
QCD scale parameter $\La$, it involves the RG invariant quark mass $\hat M$.
However, as our analysis shows, the value for $\hat m_s$ depends strongly
on $\La$: it is roughly proportional to the leading term~$L$.
We therefore present our main results in terms of
$m_s\equiv\mb_s(1\,\gev)$, which only displays a mild dependence on
$\La$. In fig.~1, we show $m_s$ and $\hat m_s$ as calculated from the
sum rule of eq.~\eqn{eq:1.7} for $\La=280$, $380$ and $480\,\mev$. The
thick lines correspond to $m_s$ and the thin lines to $\hat m_s$. Our
$\La$ corresponds to 3 flavours and to NNLO, and has been chosen such
that $\as(M_Z)=0.118\pm0.006$, which covers most values obtained from
recent analyses \cite{bet:94}. All other parameters have been set to
central values which will be discussed in the following.

Choosing the stability region from which we determine $m_s$ to be in
the range $W\equiv\sqrt u=2.0$ -- $3.0\,\gev$, from fig.~1 we obtain
$m_s=196$, 189 and $198\,\mev$ for
$\La=280$, $380$ and $480\,\mev$. The lower end of the
stability interval has been chosen such that the ${\cal O}(a)$ correction
is $\lsim50\%$ of the leading term, and at the upper end, the sum rule
becomes insensitive to the hadronic resonance structure and is
completely determined by the perturbative continuum. The parameter $s_0$
which determines the onset of the perturbative continuum has been
adjusted so as to obtain optimal stability. For the three values of $\La$
we find $s_0=5.5$, 6.4 and $7.5\,\gev^2$ respectively. It is gratifying
that $\sqrt{s_0}=2.5\pm0.2\,\gev$ is just found to be around the energy
at which the next resonance, which has not been included, is expected.
However, because of the imperfection of our phenomenological model,
we still have a residual dependence on $s_0$. This dependence is
displayed in fig.~2, where we plot $m_s$ for $\La=380\,\mev$ and
$s_0=5.9$, 6.4 and $6.9\,\gev^2$. We do not include this variation
in the error on $m_s$ because it is implicitly covered through
varying all other input parameters to be discussed below.

Let us next discuss the determination of the parameters in the
phenomenological two-point function which has been presented in
sect. 3.1. We have calculated these parameters from the experimental
data for $K\pi$-scattering given by Estabrooks et al. \cite{est:78} and
Aston et al. \cite{ast:88,awa:86}. Generally, the s-wave amplitude and
phase shift $a_S$ and $\phi_S$ can be decomposed as follows:
\begin{equation}
a_S\,e^{i\phi_S} \; = \; \frac{1}{2i}\,\biggl(\,\eta_1 e^{2i\delta_1}-1
\,\biggr)+\frac{1}{4i}\,\biggl(\,\eta_3 e^{2i\delta_3}-1\,\biggr) \,,
\label{eq:5.1}
\end{equation}
where $\eta_{1,3}$ and $\delta_{1,3}$ are the elasticities and
phase shifts for the $I=1/2$ and $I=3/2$ channel respectively.
In writing eq.~\eqn{eq:5.1} we have adopted the normalisation of
refs.~\cite{ast:88,awa:86,est:78} for $a_S$. In ref.~\cite{est:78},
$\delta_1$ has been measured for $\sqrt s=0.73$ -- $1.30\,\gev$, and
$\delta_3$ for $0.73$ -- $1.72\,\gev$.
In this work it was also demonstrated that below $1.3\,\gev$, the
scattering is purely elastic, that is, $\eta_1=\eta_3=1$. On the other
hand in ref.~\cite{awa:86} only $a_S$ and $\phi_S$ up to $\sqrt s=2.52\,\gev$
were given. In order to be able to calculate $\delta_1$ from the
data by Aston et al. above $1.3\,\gev$, we have to subtract the
$I=3/2$ contribution. We fitted this contribution to a pure
effective range
\begin{equation}
\tan\delta_3(s) \; = \; \alpha q(s)\Big[\,1+\beta q(s)^2\,\Big] \,,
\label{eq:5.2}
\end{equation}
where $q(s)$ is the centre of mass momentum defined in eq.~\eqn{eq:3.7}.
For the fit we used the full data set of ref.~\cite{est:78}, however
multiplying their error by a factor of 2 to get a $\chi^2/$d.o.f.
of order 1, and, below $1.3\,\gev$, for the data of ref.~\cite{awa:86}
we calculated $\delta_3$ from eq.~\eqn{eq:5.1} with $\eta_1=\eta_3=1$.
Our best fit is obtained for:
$\alpha=-1.04\,\gev^{-1}$ and $\beta=-0.67\,\gev^{-2}$. The value for
$\alpha$ corresponds to a scattering length of $-0.20\,fm$.

Using this fit, we have then calculated $\delta_1$
for the data of both groups below $1.7\,\gev$. The result obtained
for $\delta_1$ can be fitted to the sum of an effective range and
a Breit-Wigner resonance:
\begin{equation}
\delta_1(s) \; = \; \delta_{ER}(s) + \delta_{BW}(s) \,,
\label{eq:5.3}
\end{equation}
where $\delta_{ER}$ is again of the form given in \eqn{eq:5.2}, and
\begin{equation}
\tan\delta_{BW}(s) \; = \; \frac{M_R\Gamma_R(s)}{(M_R^2-s)} \,.
\label{eq:5.4}
\end{equation}
$\Gamma_R(s)$ has been defined in eq.~\eqn{eq:3.7}. The lowest lying
resonance in this channel is the $K_0^*(1430)$, and for our best fit
we obtain: $M_{K_0^*(1430)}= 1423\pm10\,\mev$,
$\Gamma_{K_0^*(1430)}=268\pm25\,\mev$, $\alpha=2.06\,\gev^{-1}$ and
$\beta=-1.37\,\gev^{-2}$. In this case $\alpha$ corresponds to a
scattering length of $0.41\,fm$. $M_{K_0^*(1430)}$ and
$\Gamma_{K_0^*(1430)}$ are in good agreement to the values given
in \cite{ast:88}.
We show our fit in fig.~3 together with the data points and a fit
to a pure effective range below $1.3\,\gev$. We do not give
errors for $\alpha$ and $\beta$ explicitly, because they are not
direct input parameters, but we have varied them for the calculation
of $d(s_+)$. We have also compared our fit to $\delta_1$ with other
approaches, namely the $k$-matrix formalism \cite{lp:80}, chiral
perturbation theory \cite{bkm:91,dp:93} and the quark Born diagram
formalism \cite{lgbs:94}, finding good agreement. However, the $\chi^2$
for our effective range plus Breit-Wigner Ansatz is lowest.

Above $1.7\,\gev$, we do not know how to subtract the $I=3/2$ contribution.
We therefore assume two extreme cases for our evaluation of the errors on
$d(s_+)$. As a lower bound for $\delta_1$, we use the pure effective range
also displayed in fig.~3 up to infinity, and as an upper bound, above
$1.7\,\gev$ we take $\delta_1=180^\circ$ constant \cite{nprt:83}. Using
eq.~\eqn{eq:3.8} with
\begin{equation}
d(0) \; = \; (M_K^2-M_\pi^2)\,f_+(0) \; = \; 0.22\,\gev^2 \,,
\label{eq:5.5}
\end{equation}
and varying $\alpha$ and $\beta$ within the 1 $\sigma$ level, we find
\begin{equation}
d(s_+) \; = \; 0.33\pm0.02\,\gev^2 \,.
\label{eq:5.5a}
\end{equation}
This value can be compared with other methods to calculate $d(s_+)$.
Close to $s=0$, $d(s)$ can be approximated by a linear function,
\begin{equation}
d(s) \; = \; d(0)\,\Big(\,1+\lambda_0\,\frac{s}{M_\pi^2}\,\Big) \,.
\label{eq:5.6}
\end{equation}
Using $\lambda_0=0.019$ \cite{don:74,gl:85}, we get $d(s_+)=0.31\,\gev^2$.
However, we know that $d(s)$ is a concave function of $s$ \cite{gl:85}.
Thus the latter value should lie on the low side of the true value
for $d(s_+)$. On the other hand, calculating $d(s_+)$ from chiral
perturbation theory \cite{gl:85}, we obtain $d(s_+)=0.35\,\gev^2$,
also within the errors of our determination from experiment. Nevertheless,
improved data for low energy $K\pi$-scattering would be of great help
for a more precise determination of the hadronic parameters.

For the mass and the width of the second resonance $K_0^*(1950)$ we
use the values found by Aston et al.: $M_{K_0^*(1950)}=1945\pm22\,\mev$
and $\Gamma_{K_0^*(1950)}=201\pm86\,\mev$. In addition, we need the ratio
of the decay constants for the first and second resonance which
determines their relative strength. From the dual-model vertex function,
one finds the following parametrisation \cite{dom:84},
\begin{equation}
\frac{F_2}{F_1} \; = \; -\,\gamma\,\frac{M_1^2}{M_2^2} \,.
\label{eq:5.6a}
\end{equation}
In our analysis we shall use $\gamma=0.5\pm0.3$. The lower value means
almost no contribution of the second resonance, whereas for the upper
value, the second resonance is nearly as strong as the first one. Both
being extreme cases, we think that the above range for $\gamma$ is a
conservative one. The induced uncertainty on $m_s$ from this parameter
then turns out to be $\pm\,6\,\mev$.

Taken together, the errors in all phenomenological parameters lead
to an uncertainty in the strange quark mass of $\pm\,17\,\mev$, where
we have added the errors quadratically. The dominant uncertainties
in this set of parameters are due to $d(s_+)$ and $\Gamma_{K_0^*(1430)}$,
namely $\pm\,12\,\mev$ and $\pm\,9\,\mev$ respectively. A detailed table
containing the relevant parameters entering the sum rule, their errors,
and the resulting uncertainty in $m_s$ can be found in appendix~D.

For the theoretical part of the sum rule, we still have to discuss
explicit quark mass corrections, the non-perturbative condensate
parameters and the ${\cal O}(a^3)$ correction. As it turns out,
because we work at a relatively high scale ($2.0$ -- $3.0\,\gev$),
the quark mass and condensate contributions --- except, of course,
for the global quark mass factor --- to the scalar sum rule are
negligible. For completeness we nevertheless present our input
parameters. For the light quark masses, we adopted the Particle Data values,
$m_u\equiv\overline m_u(1\,\gev)=5\,\mev$ and
$m_d\equiv\overline m_d(1\,\gev)=9\,\mev$ \cite{pdg:92}.
For the quark condensates we assume SU(3) flavour symmetry,
because anyhow their contribution is tiny, and take a standard value
\begin{equation}
\langle\bar qq\rangle(1\,\gev) \; = \; -\,(225\pm25\,\mev)^3 \,,
\label{eq:5.7}
\end{equation}
where $q=u,\,d,\,s$.
Let us, nevertheless, remark on the flavour dependence of the
quark condensate. Because we use non-normal-ordered condensates,
the strange condensate $\langle\bar ss\rangle$ is closer to the
light quark condensate $\langle\bar qq\rangle$ than had we used
normal-ordered condensates. Using eq.~\eqn{eq:c.1}, we find
$\lnp\!\bar ss\!\rnp/\langle\bar ss\rangle\approx0.88$. Together with the
known value of the flavour breaking for the normal-ordered condensate
\cite{nar:89a,cdks:81,djn:89} this leads to
$\langle\bar ss\rangle\approx0.7\langle\bar qq\rangle$.

For the gluon condensate we use an average over most recent
determinations, however with a quite conservative error,
\begin{equation}
\FF \; = \; 0.024 \pm 0.012\,\gev^4 \,.
\label{eq:5.9}
\end{equation}
This value corresponds to 1 -- 3 times the standard SVZ value \cite{svz:79}.
Finally, for the mixed condensate we take \cite{op:88}
\begin{equation}
\langle g\bar q\sigma Fq\rangle \; = \; M_0^2 \,\langle\bar qq\rangle
\qquad \hbox{with} \qquad
M_0^2 \; = \; 0.8 \pm 0.2 \, \gev^2 \,.
\label{eq:5.10}
\end{equation}
Again for the mixed condensate we can study its flavour dependence.
Using the RG invariant combination of dimension 5 \cite{nt:83,mm:93}, we
find $\lnp\!g\bar s\sigma Fs\!\rnp/\langle g\bar s\sigma Fs\rangle\approx0.99$.
Therefore, the flavour breaking for the normal- and non-normal-ordered mixed
condensates are almost equal. Using the values of refs.~\cite{op:88,bd:92},
we thus have
$\langle g\bar s\sigma Fs\rangle\approx0.75\langle g\bar q\sigma Fq\rangle$.

The last uncertainty which we have to discuss, and unfortunately the
dominant one, stems from the ${\cal O}(a^3)$ correction. At $W=2.5\,\gev$
and $\La=380\,\mev$, the ${\cal O}(a)$ correction amounts to $\approx40\%$
and the ${\cal O}(a^2)$ correction to $\approx30\%$ of the leading term.
Because the perturbative expansion has only asymptotic character, we
conclude that the uncertainty coming from the as yet unknown ${\cal O}(a^3)$
correction could be as large as 30\%. In turn this would correspond to
$c_{31}\approx580$ if the correction is positive. For the error on
$m_s$ we then find $\pm\,27\,\mev$. This uncertainty turns out to be
approximately halve of the error on the ${\cal O}(a^3)$ correction,
because the sum rule scales like $m_s^2$. Let us again emphasise that
this is not completely consistent, because $\beta_4$ and $\gamma_4$ are
unknown too, and hence have not been included. From all of the above, our
main result for $m_s$ from the scalar sum rule is:
\begin{equation}
m_s \; = \; \mb_s(1\,\gev) \; = \; 189\pm32\,\mev \,.
\label{eq:5.11}
\end{equation}
We shall, however, try to get some idea about the ${\cal O}(a^3)$
correction. From eq. \eqn{eq:2.8} it is seen that the first 3
coefficients in the perturbative expansion grow almost geometrically.
If this geometric growth continues for the ${\cal O}(a^3)$ term, we
would find $c_{31}\approx440$. This would correspond to a negative
5\% correction, leading to $m_s=178\pm18\,\mev$ if the phenomenological
error is included.

So far, we only dealt with the sum rule of eq.~\eqn{eq:1.7}. Let us now
briefly analyse the 1st moment sum rule ~\eqn{eq:1.8}. In fig.~4, we
show the 1st moment sum rule for $\La=380\,\mev$ and $s_0=5.4\,\gev^2$ in
comparison to the 0th moment sum rule at $s_0=6.4\,\gev^2$. All other
parameters take their central values. It is obvious that the 1st moment
sum rule is less stable. This is due to the fact that, because of the
additional factor $s$ in the dispersion integral, higher states are
less suppressed and the sum rule becomes more sensitive to these
corrections. Within the errors the value of $m_s$ thus obtained agrees
with our previous result. Because the 1st moment sum rule is less
stable, we refrain from taking an average with the value of eq.
\eqn{eq:5.11}, but we take it as a corroboration of this result.

\subsection{Pseudoscalar two-point function}

The hadronic input parameters for the higher resonances in the
pseudoscalar channel are less well established than in the scalar
channel. There are indications for two resonances with the same
quantum numbers as the $K$-meson, the $K(1460)$ and the $K(1830)$
\cite{pdg:92}. The reported widths are
$\Gamma_{K(1460)}=\Gamma_{K(1830)}\approx250\,\mev$, although the
error is probably large. Since we shall show in the following that
anyhow the pseudoscalar sum rule at present should be abandoned,
we just use the central values. The ratio of the decay constants
of the two resonances is again given by eq.~\eqn{eq:5.6a} taking
$\gamma=0.5$.

For the input parameters on the theoretical side of the sum rule,
we use the same set of values as for the scalar sum rule in the
previous section. In fig.~5, we display $m_s$ calculated from the
pseudoscalar sum rule for $\La=280$, $380$ and $480\,\mev$.
The values for $m_s$, obtained from this figure are $144\,\mev$,
$136\,\mev$ and $143\,\mev$ for $s_0=4.9\,\gev$, $5.9\,\gev$ and
$6.9\,\gev$, and the three values of $\La$, respectively.
Even though the errors on the ${\cal O}(a^3)$ correction are large,
we would expect that the central values for $m_s$ from scalar and
pseudoscalar sum rules agree within the errors of the hadronic
parametrisation, since the radiative corrections for flavour non-singlet
scalar and pseudoscalar two-point functions are the same.
This is, however, not the case. Let us comment on the origin of this
discrepancy.

Because we evaluate the sum rule at a rather high scale, 2 -- $3\,\gev$,
the dominant hadronic contribution comes from the $K(1430)$. Therefore,
the error on the global coefficient of this resonance determines the
largest error from the hadronic spectral function. In the case of the
scalar sum rule, the corresponding quantity was the scalar form factor
at threshold, $d(s_+)$. Had we used instead the leading order chiral
perturbation theory result, $d(s_+)=d(0)=0.22\,\gev^2$, the central value
of $m_s$ would have been $127\,\mev$ --- $m_s$ scales linearly
with $d(s_+)$! The corrections from NLO chiral perturbation theory were
large and led to $d(s_+)=0.35\,\gev^2$, quite close to the value
determined from experiment.

For the pseudoscalar spectral function, we neither have available a
NLO chiral perturbation theory calculation of the $(K\pi\pi)$-contribution,
nor is there enough data in this system to calculate the corresponding
pseudoscalar form factor from experiment. Nevertheless, an additional
hint that the higher order corrections are large comes from a NLO
chiral perturbation theory calculation of the matrix elements in
$K_{l4}$-decays \cite{bcg:94}, which are related to the matrix elements
of eqs.~\eqn{eq:3.10} and \eqn{eq:3.11}. We shall return to a calculation
of the pseudoscalar spectral function at NLO chiral perturbation theory
in a forthcoming publication.

A different possible parametrisation of the pseudoscalar spectral
function is the zero-width approximation also for the higher resonances,
\begin{equation}
\rho_5(s) \; = \; F_K^2 M_K^4\,\biggl\{\,\delta(s-M_K^2) + r_K
\delta(s-M_K'^2) + r_K\gamma^2 \delta(s-M_K''^2)\,\biggr\} \,,
\label{eq:5.13}
\end{equation}
where
\begin{equation}
r_K \; \equiv \; \frac{F_{K'}^2 M_{K'}^4}{F_K^2 M_K^4}
\; = \; 7\pm1
\label{eq:5.14}
\end{equation}
was also determined from a sum rule analysis \cite{nar:89a}. Here we adopted
the abbreviations $K'\equiv K(1460)$ and $K''\equiv K(1830)$. If we insert
this parametrisation of the spectral function into the sum rule of
eq.~\eqn{eq:1.7}, we find as a central value $m_s=162\,\mev$, on the
low side of the strange mass region as calculated from the scalar sum
rule. Let us point out, however, that the zero-width approximation only
serves as a lower bound on the spectral function, and the corrections
are expected to be of the order of 10 -- 20\% \cite{dom:84}. In addition,
it is unclear whether the error on $r_K$ given in eq.~\eqn{eq:5.14}
is a conservative one.

To conclude, due to unknown uncertainties in the hadronic parametrisation
of the pseudoscalar spectral function, at present, we shall not use the
analysis as an independent determination of $m_s$, but we plan to return
to this issue in a future publication.

\newsection{Discussion}

Before we summarise our main results, let us discuss a few previous
determinations of $m_s$ which are of interest in view of the present
analysis.

In 1982, Gasser and Leutwyler \cite{gl:82} reviewed the status of
quark mass determinations known at that time. For the determination
of $m_s$, they used a combination of different QCD sum rules, together
with quark mass ratios from chiral perturbation theory. Their final
quoted value was $m_s=180\pm50\,\mev$. A calculation using finite
energy sum rules at NNLO for scalar and pseudoscalar two-point
functions was performed by Gorishny et al. \cite{gkl:84}. For the
hadronic parametrisation of the two-point functions they assumed
the zero-width approximation and found $m_s=195\,\mev$, where the
error was estimated to be of the order of 25\%.

The hadronic parametrisation of the two-point function in the
scalar channel was improved in refs.~\cite{nprt:83,dom:84,dl:85}
by taking resonances with a finite width. Whereas the authors of
\cite{nprt:83} only gave a lower bound on the strange quark mass,
the final result of ref.~\cite{dl:85} was $m_s=192\pm15\,\mev$. Since
this was only a NLO calculation, due to the large radiative corrections,
the error in this determination appears to be too small.
In ref.~\cite{dr:87}, Dominguez and de Rafael determined the
strange mass by first calculating $m_u+m_d$ from the pseudoscalar
sum rule for light quarks and then applying quark mass ratios from
chiral perturbation theory to find $m_s=199\pm33\,\mev$. They also
refrained from using the pseudoscalar sum rule for the $K$-channel
directly because of large uncertainties in the hadronic parametrisation.
Although it is not obvious that the parametrisation of the spectral
function is more reliable in the case of light quarks $m_u$ and
$m_d$, their central value is in good agreement to our results.
The work of Narison is summarised in refs.~\cite{nar:89a,nar:89}.
By purely analysing the pseudoscalar channel, he obtained
$m_s=169\pm10\,\mev$, whereas his final result for an average over
determinations from different sum-rule channels gave $m_s=160\pm9\,\mev$.
The further reduction of $m_s$ is caused by the rather low value
found in the $\phi\,$-meson channel, although the errors are very large.

The last sum-rule
determination which we would like to discuss, and the one which
comes closest to our analysis of the scalar channel, is the work
by Dominguez et al.~\cite{dgp:91}. The main differences are the
following: they only used NLO expressions for the perturbative
two-point function, they did not include the cancellation of mass
singularities for the gluon condensate and the hadronic spectral
function had been parametrised differently. Nevertheless,
their final result, $m_s=194\pm4\,\mev$, is in good agreement to
our findings. However, the quoted error was obtained by
varying solely $\La$. Since this is a minor source of uncertainty,
to our minds, their error is largely underestimated. Very recently,
the strange quark mass was calculated from lattice QCD \cite{all:94}.
The result was $\mb_s(2\,\gev)=127\pm18\,\mev$. Scaling this value
down to $\mu=1\,\gev$, we find $m_s=180\pm25\,\mev$.

All determinations of the strange quark mass from QCD sum rules
which have been discussed above are in general agreement
with the calculation presented in this work, although in some analyses
the errors were underestimated. Our main result was obtained from
a sum rule for the divergence of the vector current:
$m_s=189\pm32\,\mev$. In this determination, the error is dominated
by the unknown perturbative ${\cal O}(a^3)$ correction. We tried to
estimate the ${\cal O}(a^3)$ correction by observing a geometric growth
of the first three terms in the perturbation series, and assuming
that this geometric behaviour continues for the next order. The
result then is: $m_s=178\pm18\,\mev$, where the error now dominantly
stems from the hadronic parametrisation of the spectral function.
Especially this latter value is in astonishing agreement with the lattice
calculation of ref.~\cite{all:94}. The sum rule for the divergence of
the axial-vector was found to be unreliable, due to large uncertainties
in the hadronic parametrisation of the spectral function.
Recently, instanton contributions to the determination of light quark
masses were investigated \cite{gn:93}, and it was found that they can
be sizeable if the scale at which the sum rule is evaluated, is too low.
Because of the relatively large scale at which we obtain the strange
mass, in the case at hand these contributions can be safely neglected.

As a final application of our results, we estimate the masses of the
light quarks $m_u$ and $m_d$ from quark-mass ratios calculated in
chiral perturbation theory, and the mass of the strange quark obtained
in the present work. The relevant quark-mass ratios are \cite{gl:84,gl:85}
\begin{equation}
\frac{2m_s}{m_u+m_d} \; = \; 25.7\pm2.6
\qquad \mbox{and} \qquad
\frac{m_d-m_u}{m_u+m_d} \; = \; 0.28\pm0.03 \,.
\label{eq:6.1}
\end{equation}
Using $m_s=189\pm32\,\mev$, we find $m_u=5.3\pm1.5\,\mev$ and
$m_d=9.4\pm1.5\,\mev$, whereas for $m_s=178\pm18\,\mev$ we obtain
$m_u=5.0\pm1.1\,\mev$ as well as $m_d=8.9\pm1.1\,\mev$, in good
agreement to earlier determinations published in the literature.
We plan to directly investigate quark-mass ratios from QCD sum
rules in the future.

\vskip 1cm \noindent
{\Large\bf Acknowledgement}

\noindent
It is a pleasure to thank A. J. Buras, H. G. Dosch and A. Pich
for helpful discussions. M.~J. would also like to thank T. Barnes
and E. de Rafael for e-mail discussions, C. Ayala and J. R. Pel{\'a}ez
for providing Fortran programs related to their work and
D. Pirjol for making available a preliminary version of a paper
on the same subject, prior to publication.

\newpage
\noindent

\appendix{\LARGE\bf Appendices}
\newsection{The Borel transform}

The Borel operator ${\cal B}$ is defined by ($s\equiv Q^2$)
\begin{equation}
{\cal B}_u \; \equiv \; \lim_{\stackrel{s,n\to\infty}{s/n=u}}
\frac{(-s)^n}{(n-1)!}\,\frac{\partial^{\,n}}{(\partial s)^n} \,.
\label{eq:a.1}
\end{equation}
The Borel transformation is an inverse Laplace transform \cite{wid:46}.
If we set
\begin{equation}
\wh f(u) \; \equiv \; {\cal B}_u\Big[\,f(s)\,\Big] \,,
\qquad \hbox{then} \qquad
f(s) \; = \; \int\limits_0^\infty \frac{1}{u}\,\wh f(u)\,e^{-s/u}du \,.
\label{eq:a.2}
\end{equation}
We need the following Borel transforms \cite{pbm:92}:
\begin{equation}
{\cal B}_u\biggl[\,\frac{1}{s^\alpha}\,\ln^n\frac{s}{\mu^2}\,\biggr] \; = \;
\frac{1}{u^\alpha}\sum_{k=0}^n (-)^{n-k}{n\choose k}\ln^k\frac{u}{\mu^2}\,
\Biggl[\,\frac{\partial^{\,n-k}}{(\partial\alpha)^{n-k}}\,\frac{1}
{\Gamma(\alpha)}\,\Biggr] \,,
\label{eq:a.3}
\end{equation}
and explicitly for $n=0,\,1,\,2,\,3$:
\begin{eqnarray}
\label{eq:a.4}
{\cal B}_u\biggl[\,\frac{1}{s^\alpha}\,\biggr] & = &
\frac{1}{u^\alpha \Gamma(\alpha)} \,, \nn \\
\smvs
{\cal B}_u\biggl[\,\frac{1}{s^\alpha}\,\ln\frac{s}{\mu^2}\,\biggr] & = &
\frac{1}{u^\alpha \Gamma(\alpha)}\,\Big[\,\bar L+\psi(\alpha)\,\Big] \,, \\
\smvs
{\cal B}_u\biggl[\,\frac{1}{s^\alpha}\,\ln^2\frac{s}{\mu^2}\,\biggr] & = &
\frac{1}{u^\alpha \Gamma(\alpha)}\,\Big[\,\bar L^2+2\psi(\alpha)\bar L+
\psi^2(\alpha)-\psi'(\alpha)\,\Big] \,,\nn \\
\smvs
{\cal B}_u\biggl[\,\frac{1}{s^\alpha}\,\ln^3\frac{s}{\mu^2}\,\biggr] & = &
\frac{1}{u^\alpha \Gamma(\alpha)}\,\Big[\,\bar L^3+3\psi(\alpha)\bar L^2+
3\big(\psi^2(\alpha)-\psi'(\alpha)\big)\bar L+\psi^3(\alpha) \nn \\
\tvs
& & \hspace{22mm} -\,3\psi(\alpha)\psi'(\alpha)+\psi''(\alpha)\,\Big] \,. \nn
\end{eqnarray}
Here, $\bar L\equiv\ln u/\mu^2$ and
\begin{equation}
\psi^{(k)}(\alpha) \; \equiv \; \frac{\partial^{\,k+1}}{(\partial\alpha)^{k+1}}
\,\ln\Gamma(\alpha) \,.
\label{eq:a.5}
\end{equation}
For integer values of $\alpha$, we have the following useful
formulae \cite{gr:80}:
\begin{eqnarray}
\psi(n) & = & \sum_{l=1}^{n-1}\frac{1}{l}-\gl \,, \nn \\
\smvs
\psi^{(k)}(n) & = & (-)^k\,k!\,\Biggl[\,\sum_{l=1}^{n-1}
\frac{1}{l^{k+1}}-\zeta(k+1)\,\Biggr] \,;\qquad k\ge1 \,.
\label{eq:a.6}
\end{eqnarray}
To arrive at eq.~\eqn{eq:1.5}, we also need:
\begin{equation}
{\cal B}_u\biggl[\,\frac{1}{(x+s)^\alpha}\,\biggr] \; = \;
\frac{1}{u^\alpha \Gamma(\alpha)}\,e^{-x/u} \,.
\label{eq:a.7}
\end{equation}

\section{Renormalization group functions}

For the definition of the renormalization group functions we
follow the notation of Pascual and Tarrach \cite{pt:84}, except
that we defined the $\beta$-function such that $\beta_1$ is positive.
The expansions of $\beta(a)$ and $\gamma(a)$ take the form:
\begin{equation}
\beta(a) \; = \; -\,\beta_1 a-\beta_2 a^2-\beta_3 a^3-\ldots \,,
\quad \hbox{and} \quad
\gamma(a) \; = \; \gamma_1 a+\gamma_2 a^2+\gamma_3 a^3+\ldots \,,
\label{eq:b.1}
\end{equation}
with
\begin{equation}
\beta_1 \; = \; \frac{1}{6}\,\Big[\,11N-2f\,\Big] \,, \qquad
\beta_2 \; = \; \frac{1}{12}\,\Big[\,17N^2-5Nf-3C_Ff\,\Big] \,,\qquad\qquad {}
\label{eq:b.2}
\end{equation}

\begin{displaymath}
\beta_3 \; = \; \frac{1}{32}\,\biggl[\,\frac{2857}{54}N^3-\frac{1415}{54}N^2f+
\frac{79}{54}Nf^2-\frac{205}{18}NC_Ff+\frac{11}{9}C_Ff^2+C_F^2f\,\biggr] \,,
\label{eq:b.3}
\end{displaymath}
and
\begin{equation}
\gamma_1 \; = \; \frac{3}{2}\,C_F \,, \qquad\qquad\qquad
\gamma_2 \; = \; \frac{C_F}{48}\,\Big[\,97N+9C_F-10f\,\Big] \,,
\hspace{2cm}{} \nn
\label{eq:b.4}
\end{equation}

\begin{displaymath}
\gamma_3 \; = \; \frac{C_F}{32}\,\biggl[\,\frac{11413}{108}N^2-\frac{129}{4}
NC_F-\Big(\frac{278}{27}+24\ze\Big)Nf+\frac{129}{2}C_F^2-\Big(23-24\ze\Big)
C_Ff-\frac{35}{27}f^2\,\biggr] \,,
\label{eq:b.5}
\end{displaymath}
where $\gamma_3$ has been taken from ref. \cite{tar:82}.

We also need the expansion of an arbitrary product $\ab^m(u)\Mb^{\,n}(u)$
in inverse powers of $L\equiv\ln(u/\Lambda^2)$:
\begin{equation}
\ab^m(u)\,\Mb^{\,n}(u) \; = \; \frac{\hat M^n}{\beta_1^m}\,\biggl(\frac{2}{L}
\biggr)^{m+n\,\gamma_1/\beta_1}\biggr[\,1+d_{11}\,\frac{\ln L}{L} +
\frac{d_{10}}{L} + d_{22}\,\frac{\ln^2 L}{L^2} + d_{21}\;\frac{\ln L}{L^2} +
\frac{d_{20}}{L^2}\,\biggr] \,,
\label{eq:b.6}
\end{equation}
with
\begin{eqnarray}
d_{11} & = & -\,\frac{2\beta_2}{\beta_1^3}\,\Big(m\,\beta_1+n\,\gamma_1
\Big) \,, \hspace{1cm}
d_{10} \; = \; \frac{2n}{\beta_1^3}\,\Big(\beta_1\gamma_2-\beta_2\gamma_1
\Big) \,,\nn \\ \nn \\
d_{22} & = & \frac{1}{2}\,d_{11}^2-\frac{\beta_2}{\beta_1^2}\,d_{11} \,,
\hspace{2.2cm}
d_{21} \; = \; d_{11} d_{10} - \frac{4\beta_2}{\beta_1^4}\,\Big(m\,\beta_2+
n\,\gamma_2\Big) \,, \\ \nn \\
d_{20} & = & \frac{1}{2}\,\Big(d_{21}-d_{11}d_{10}+d_{10}^2\Big) +
\frac{2}{\beta_1^3}\,\Big(m\,\beta_3+n\,\gamma_3\Big) - \frac{d_{11}}{
\beta_1^2\beta_2}\,\Big(\beta_1\beta_3-\beta_2^2\Big) \,.\nn
\end{eqnarray}
Here, $\hat M$ is the renormalization group invariant quark mass.
Because $\gamma_1/\beta_1$ is close to $1/2$, from the prefactor
of eq.~\eqn{eq:b.6} we read off that two powers of $M$ count
approximately like $a$ in the perturbative expansion. This way of
counting orders in perturbation theory has been adopted in this work.

Using the renormalization group equation \eqn{eq:2.5}, we find the
following relations between the $c_{ij}$:
\begin{eqnarray}
\label{eq:b.8}
c_{12} & = & -\,\frac{1}{2}\,\gamma_1 c_{01} \,, \nn \\
\tvs
c_{22} & = & -\,\frac{1}{4}\,\Big[\,\big(\beta_1+2\gamma_1\big)c_{11}+
2\gamma_2 c_{01}\,\Big] \,, \\
\tvs
c_{23} & = & \phantom{-}\,\frac{1}{12}\,\gamma_1\big(\beta_1+2\gamma_1\big)
c_{01} \,, \nn \\
\smvs
\label{eq:b.9}
c_{32} & = & -\,\frac{1}{4}\,\Big[\,2\big(\beta_1+\gamma_1\big)c_{21}+
\big(\beta_2+2\gamma_2\big)c_{11}+2\gamma_3 c_{01}\,\Big] \,, \nn \\
\tvs
c_{33} & = & -\,\frac{1}{6}\,\Big[\,(\beta_2+2\gamma_2) c_{12}+
2(\beta_1+\gamma_1) c_{22}\,\Big] \,, \\
\tvs
c_{34} & = & -\,\frac{1}{4}\,(\beta_1+\gamma_1) c_{23} \,. \nn
\end{eqnarray}

\newsection{Renormalization group invariant condensates}

To the order considered in this work the renormalization group invariant
condensates are given by~\cite{nt:83,sc:88}:
\begin{eqnarray}
\label{eq:c.1}
\Omega^{m_iq_j}_3 & \equiv & m_i\langle\bar q_j q_j\rangle + w_1\,
\biggl[\,\frac{1}{a}+w_2\,\biggr]\,m_i\, m_j^3 \,, \\
\smvs
\label{eq:c.2}
\Omega_4 & \equiv & \langle aFF\rangle - 4\,\frac{\gamma_1}{\beta_1}\,a
\sum_i m_i\langle\bar q_i q_i\rangle - \frac{\gamma_1^0}{\beta_1}
\sum_i m_i^4 \,,
\end{eqnarray}
where
\begin{equation}
w_1 \; = \; \frac{\gamma_1^0}{(4\gamma_1-\beta_1)} \,, \qquad
w_2 \; = \; \frac{1}{6}\,(4\gamma_1-\beta_1) - \frac{1}{4\gamma_1}\,
(4\gamma_2-\beta_2) \,,
\label{eq:c.3}
\end{equation}
and $\gamma_1^0=N/(2\pi^2)$ is the leading order anomalous dimension
of the vacuum energy.

\newpage
\newsection{Input parameters and the error on $m_s$}

We have omitted those input parameters from the table whose
variation results in an error on $m_s$ less than $1\,\mev$.

%
\begin{table}[thb]
\begin{center}
\renewcommand{\arraystretch}{1.4}
\begin{tabular}{|c|c|c|}
\hline
Parameter & Value & $\Delta m_s$ \\
\hline
$\La$ & $380\pm100\,\mev$ & $\pm\,5\,\mev$
\\
${\cal O}(a^3)$ & & $\pm\,27\,\mev$
\\
$d(s_+)$ & $0.33\pm0.02\,\gev^2$ & $\pm\,12\,\mev$
\\
$M_{K_0^*(1430)}$ & $1423\pm10\,\mev$ & $\pm\,3\,\mev$
\\
$\Gamma_{K_0^*(1430)}$ & $268\pm25\,\mev$ & $\pm\,9\,\mev$
\\
$\Gamma_{K_0^*(1950)}$ & $201\pm86\,\mev$ & $\pm\,4\,\mev$
\\
$\gamma$ & $0.5\pm0.3$ & $\pm\,6\,\mev$
\\
\hline
\end{tabular}
\renewcommand{\arraystretch}{1}
\end{center}
\end{table}

\newpage


\clearpage

\begin{figure}[h]
\vspace{0.1in}
\centerline{
\rotate[r]{
\epsfysize=4.5in
\epsffile{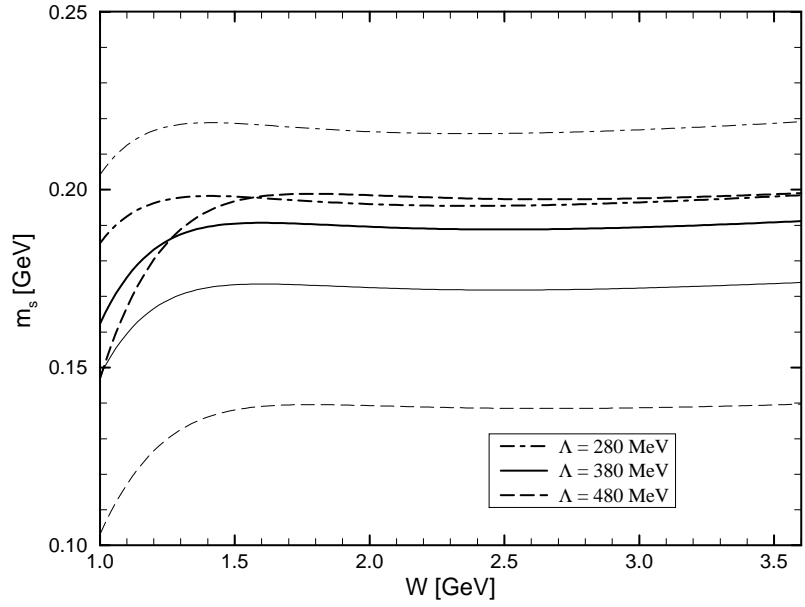}
\vspace{.1in}
}}
\caption[]{$m_s$ and $\hat m_s$ for three different values of $\La$.
Thick lines correspond to $m_s$ and thin lines to $\hat m_s$.\label{fig:1}}
\end{figure}

\begin{figure}[h]
\vspace{0.1in}
\centerline{
\rotate[r]{
\epsfysize=4.5in
\epsffile{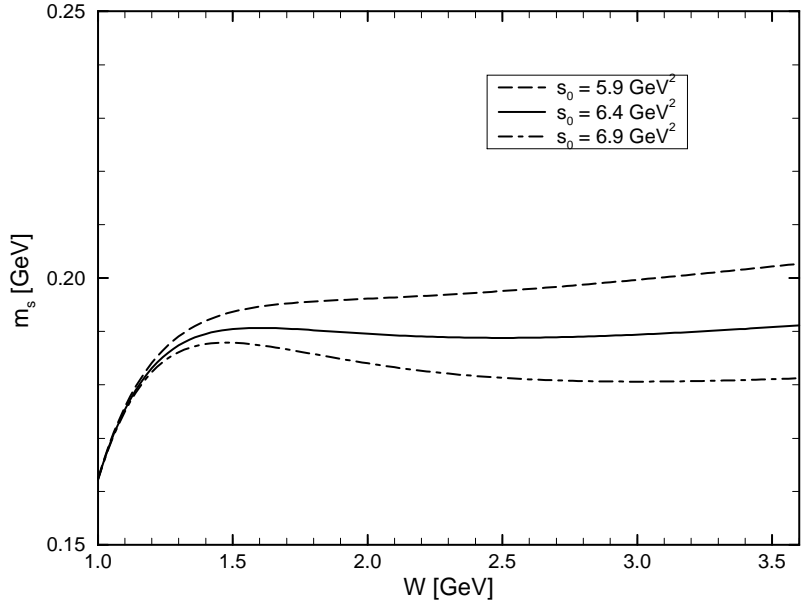}
\vspace{.1in}
}}
\caption[]{$m_s$ for three different values of $s_0$ and $\La=380\,\mev$.
\label{fig:2}}
\end{figure}

\clearpage

\begin{figure}[h]
\vspace{0.1in}
\centerline{
\rotate[r]{
\epsfysize=4.5in
\epsffile{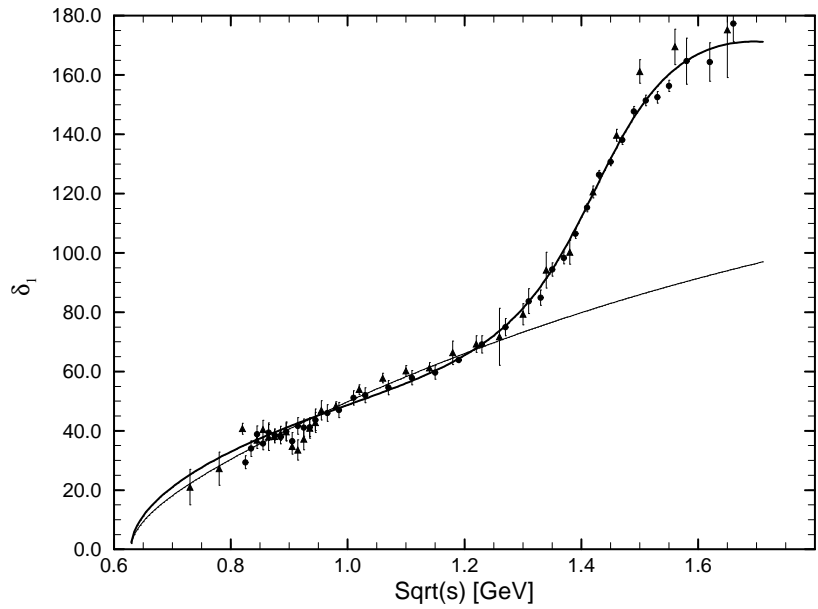}
\vspace{.1in}
}}
\caption[]{Our fit for the $s$-wave $I=1/2$ phase shift, compared to
the experimental data. The thin line corresponds to a pure effective
range.\label{fig:3}}
\end{figure}

\begin{figure}[h]
\vspace{0.1in}
\centerline{
\rotate[r]{
\epsfysize=4.5in
\epsffile{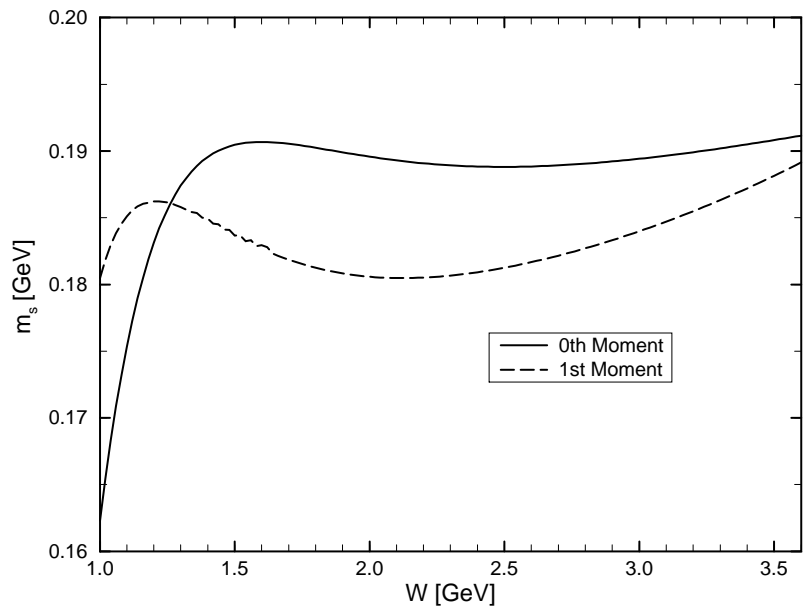}
\vspace{.1in}
}}
\caption[]{Comparison of $m_s$ from the 0th and 1st-moment sum rules.
\label{fig:4}}
\end{figure}

\clearpage

\begin{figure}[h]
\vspace{0.1in}
\centerline{
\rotate[r]{
\epsfysize=4.5in
\epsffile{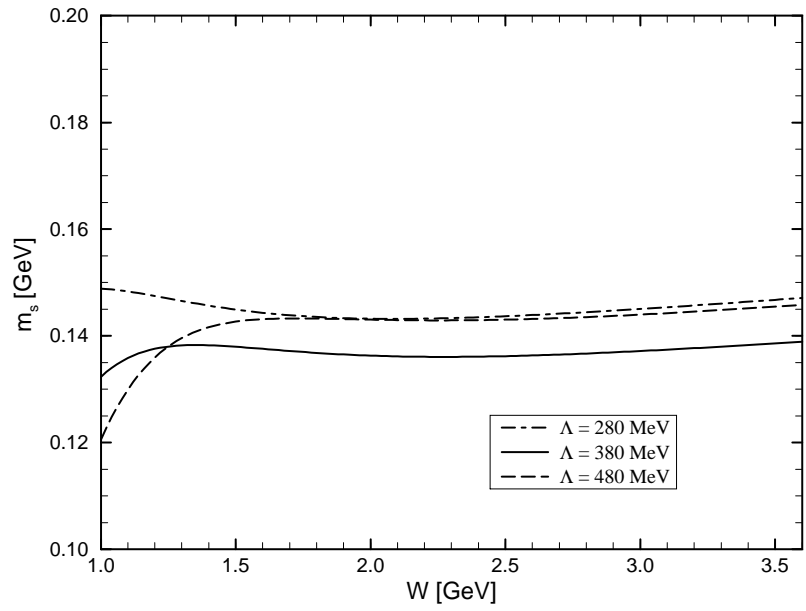}
\vspace{.1in}
}}
\caption[]{$m_s$ calculated from the pseudoscalar sum rule.\label{fig:5}}
\end{figure}

\end{document}